\documentclass[journal]{vgtc}                

\ifpdf
  \pdfoutput=1\relax                   
  \usepackage{graphicx}                
  \DeclareGraphicsExtensions{.pdf,.png,.jpg,.jpeg} 
\else
  \usepackage{graphicx}                
  \DeclareGraphicsExtensions{.eps}     
\fi%

\graphicspath{{figures/}{pictures/}{images/}{./}} 

\usepackage{microtype}                 
\PassOptionsToPackage{warn}{textcomp}  
\usepackage{textcomp}                  
\usepackage{mathptmx}                  
\usepackage{times}                     
\usepackage[usenames, dvipsnames]{color}
\usepackage[usenames,dvipsnames,table]{xcolor}
\usepackage{cite}                      
\usepackage{tabu}                      
\usepackage{booktabs}                  

\usepackage{xspace}
\usepackage{dblfloatfix}

\usepackage{cleveref}                 
\usepackage{tabularx}
\usepackage{array}
\usepackage{makecell}
\usepackage{caption}

\onlineid{0}

\newcommand{\bstart}[1]{\vspace{1mm} \noindent{\textbf{#1:}}}
\newcommand{\bpstart}[1]{\vspace{1mm} \noindent{\textbf{#1.}}}

\newcommand{\etal}{et al.\xspace}
\newcommand{\eg}{e.g., }

\vgtccategory{Research}
\vgtcpapertype{theory/model}

\title{Critical Reflections on Visualization Authoring Systems}

\author{Arvind Satyanarayan, Bongshin Lee, Donghao Ren, Jeffrey Heer,\\ John Stasko, John Thompson, Matthew Brehmer, and Zhicheng Liu}
\authorfooter{
\item
 Arvind Satyanarayan is with the Massachusetts Institute of Technology. E-mail: arvindsatya@mit.edu.
\item
 Bongshin Lee is with Microsoft Research. E-mail: bongshin@microsoft.com.
\item
 Donghao Ren is with the University of California, Santa Barbara. E-mail: donghaoren@cs.ucsb.edu.
\item
 Jeffrey Heer is with the University of Washington. E-mail: jheer@uw.edu.
\item
 John Stasko and John Thompson are with the Georgia Institute of Technology. E-mails: john.stasko@cc.gatech.edu, jrthompson@gatech.edu.
\item
 Matthew Brehmer is an independent researcher; he conducted this work while with Microsoft Research. E-mail: mb@mattbrehmer.ca.
\item
 Zhicheng Liu is with Adobe Research. E-mail: leoli@adobe.com.
}

\shortauthortitle{Satyanarayan \MakeLowercase{\textit{et al.}}: Expressivity, Learnability, \& Reusability: Critical Reflections of Visualization Authoring Systems}

\abstract{An emerging generation of visualization authoring systems support expressive information visualization without textual programming. As they vary in their visualization models, system architectures, and user interfaces, it is challenging to directly compare these systems using traditional evaluative methods. Recognizing the value of contextualizing our decisions in the broader design space, we present critical reflections on three systems we developed\,---\,Lyra, Data Illustrator, and Charticulator.
This paper surfaces knowledge that would have been daunting within the constituent papers of these three systems.
We compare and contrast their (previously unmentioned) limitations and trade-offs between expressivity and learnability.
We also reflect on common assumptions that we made during the development of our systems, thereby informing future research directions in visualization authoring systems.}


\CCScatlist{ 
 \CCScat{K.6.1}{Management of Computing and Information Systems}%
{Project and People Management}{Life Cycle};
 \CCScat{K.7.m}{The Computing Profession}{Miscellaneous}{Ethics}
}

\vgtcinsertpkg

\begin{document}

\newcommand{\figureRedBlueAmerica}{
  \begin{figure}[t!]
  \centering
  \includegraphics[width=\columnwidth]{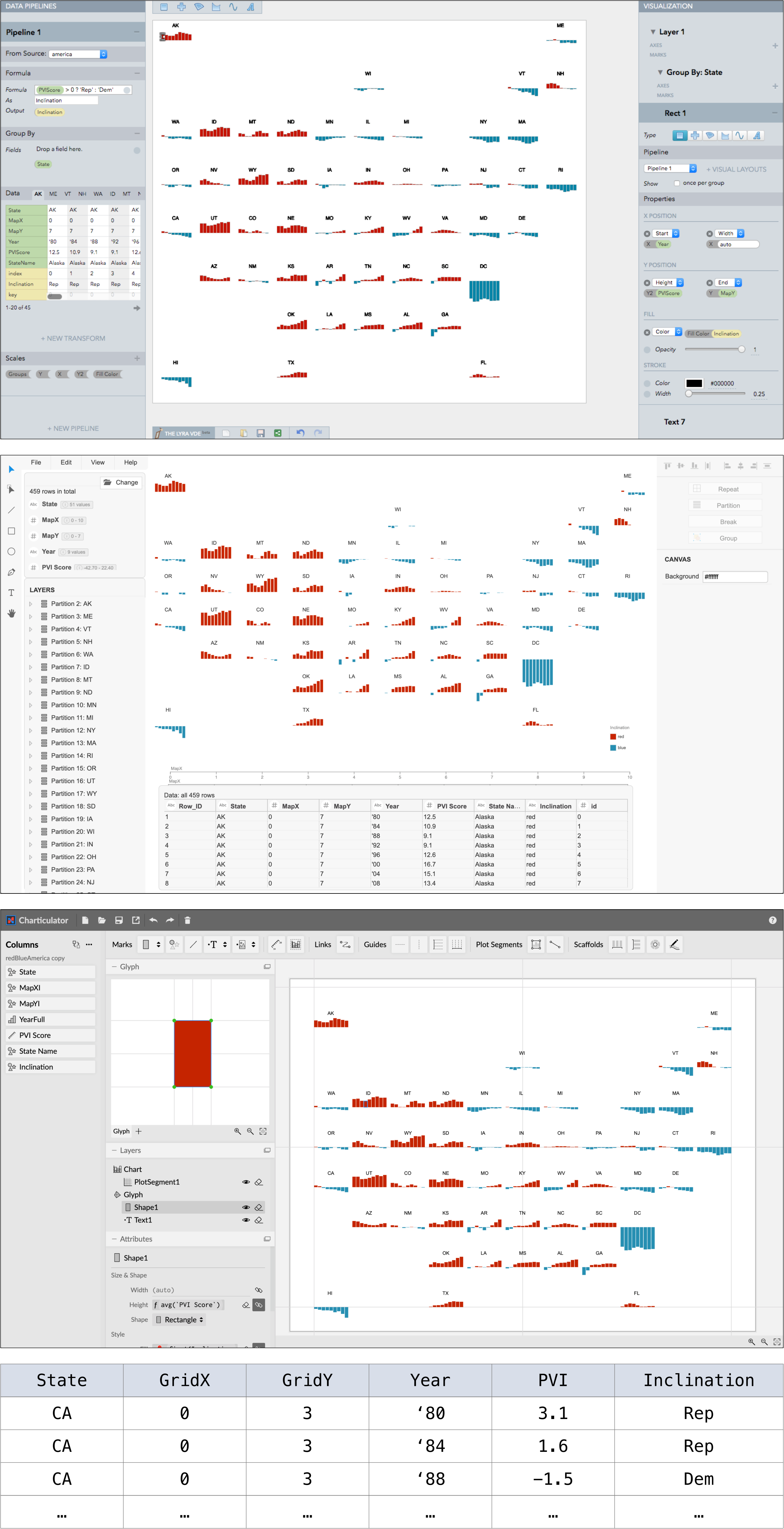}
  \vspace{-3mm}
  \caption{Recreating the Wall Street Journal's \emph{A Field Guide to Red and Blue America}~\cite{wsj:redblueamerica} visualization using Lyra, Data Illustrator, and Charticulator (top to bottom), with an excerpt of the backing dataset.}
  \vspace{-7mm}
  \label{fig:RedBlueAmerica}
  \end{figure}
}

\newcommand{\figureRedBlueAmericaNoSystem}{
  \begin{figure}[t!]
  \centering
  \includegraphics[width=\columnwidth]{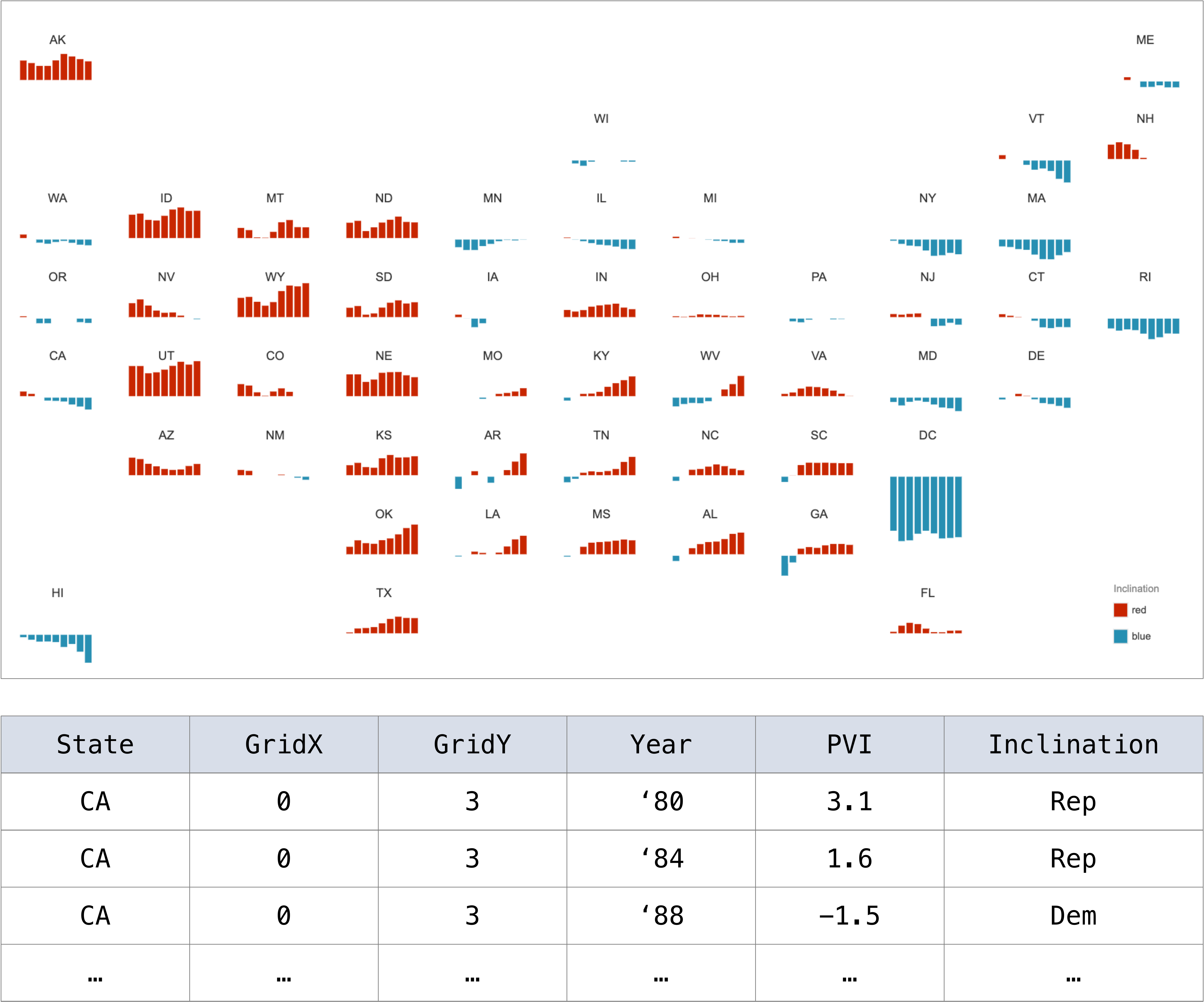}
  \caption{The Wall Street Journal's \emph{A Field Guide to Red and Blue America}~\cite{wsj:redblueamerica} visualization, with an excerpt of the backing dataset.}
  \label{fig:RedBlueAmerica}
  \vspace{-5mm}
  \end{figure}
}

\newcommand{\figureDataBinding}{
  \begin{figure*}[b!]
  \centering
  \includegraphics[width=\textwidth]{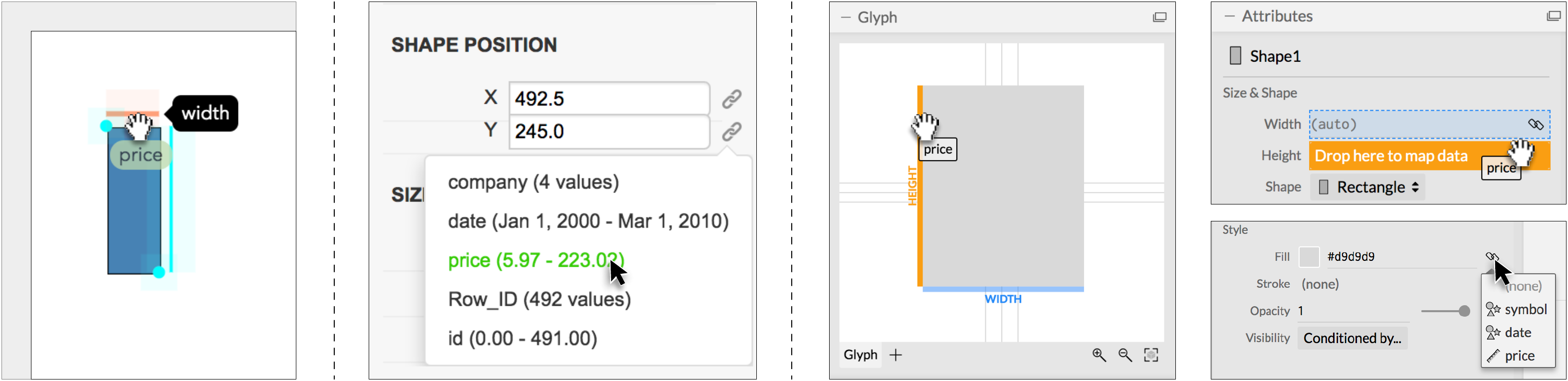}
  \vspace{-2mm}
  \caption{Data binding via dropzones in Lyra (left), via the binding icon in Data Illustrator (middle), and via either approach in Charticulator (right).}
  \label{fig:DataBinding}
  \end{figure*}
}

\newcommand{\figureScalesAxesLegends}{
  \begin{figure}[t!]
  \centering
  \includegraphics[width=\columnwidth]{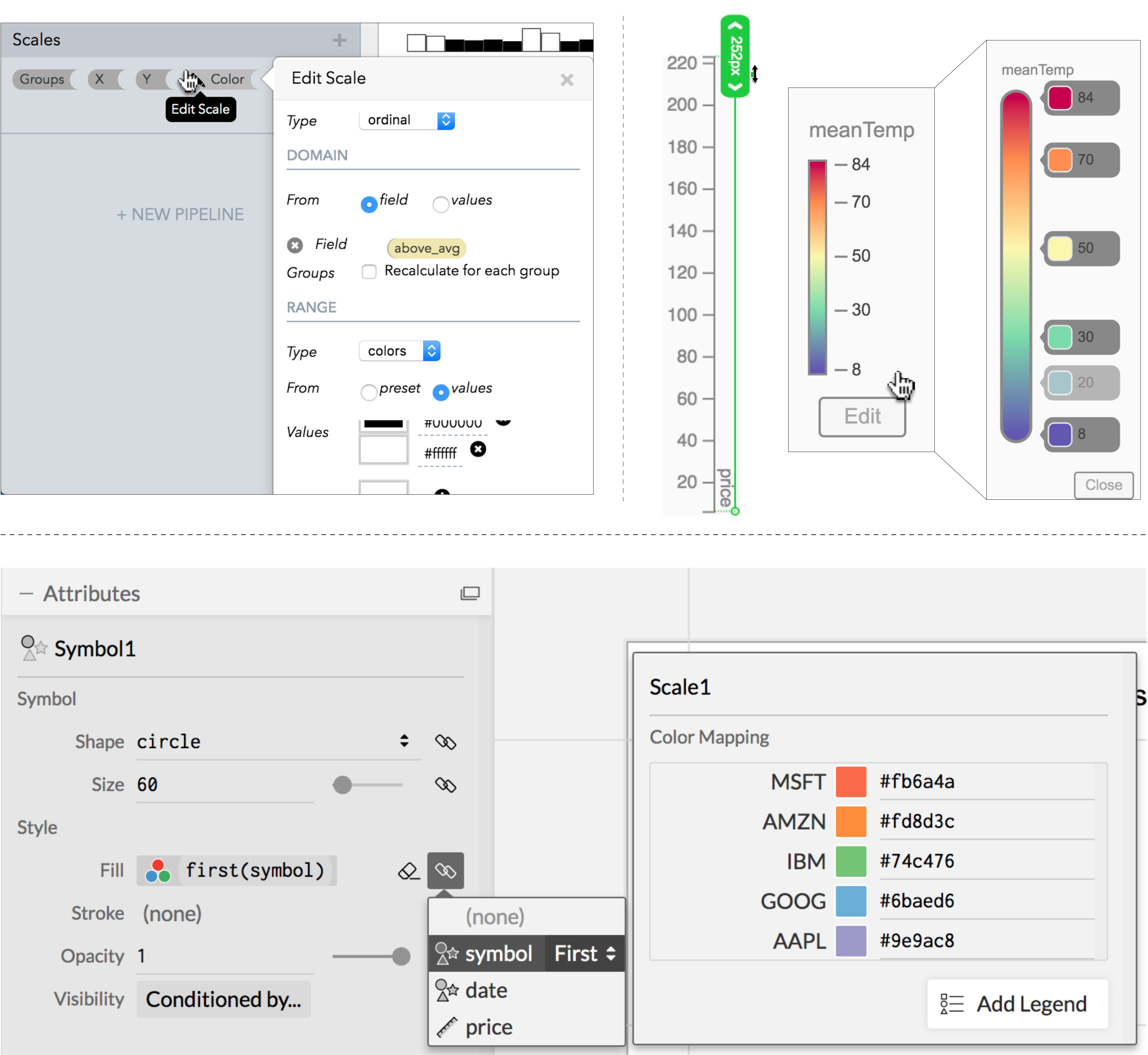}
  \caption{Lyra's scale listing and configuration panel (top-left), Data Illustrator's direct manipulation controls for axes and legends (top-right), and Charticulator's scale configuration panel (bottom).}
  \vspace{-5mm}
  \label{fig:ScalesAxesLegends}
  \end{figure}
}

\newcommand{\figureRedBlueAmericaSteps}{
  \begin{figure}[t!]
  \centering
  \includegraphics[width=\columnwidth]{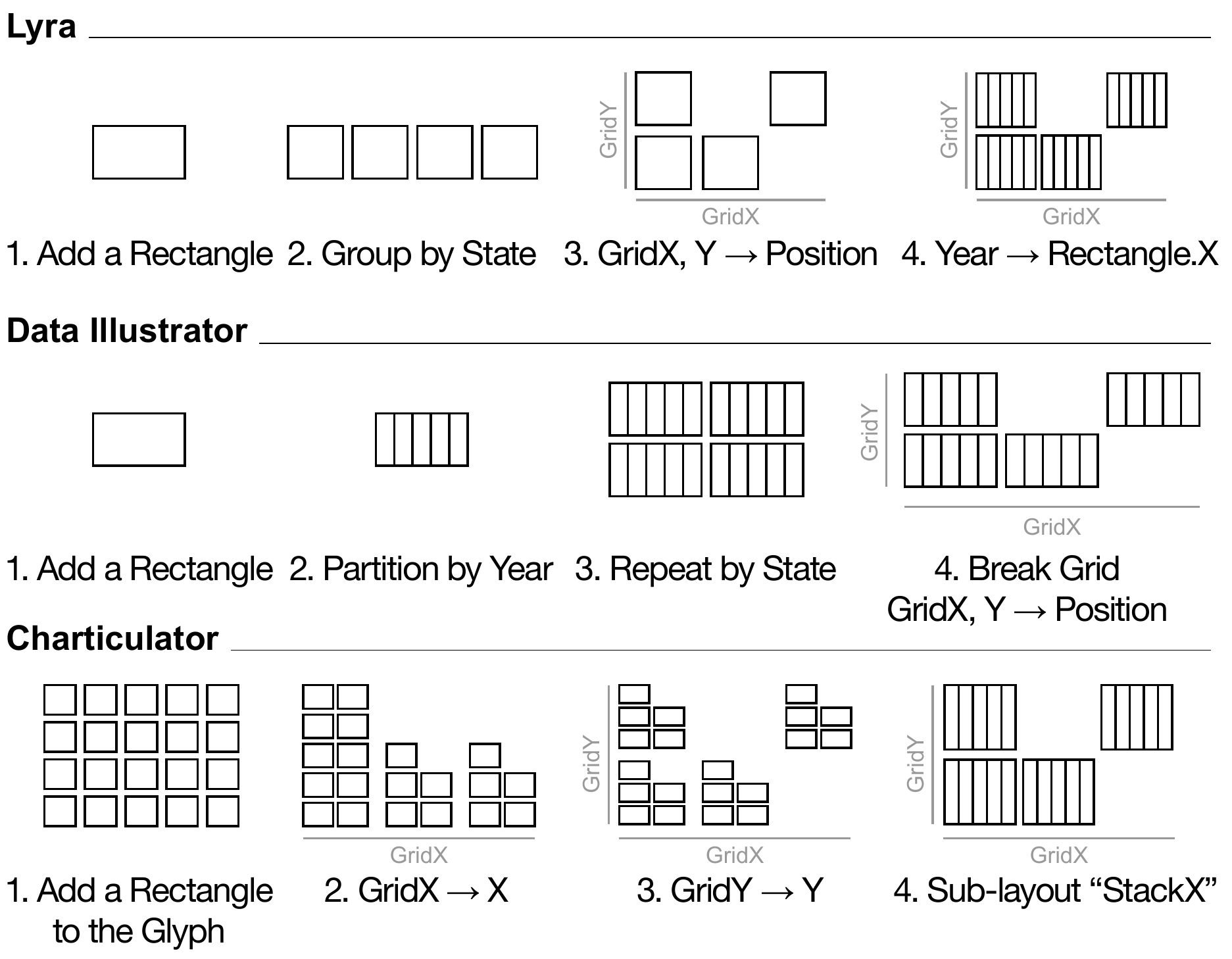}
  \caption{A step-by-step illustration of how to recreate the layout in The Wall Street Journal's \emph{A Field Guide to Red and Blue America}~\cite{wsj:redblueamerica} with each system (data is reduced to four states and five years for simplicity).}
  \label{fig:RedBlueAmericaSteps}
  \vspace{-3mm}
  \end{figure}
}


\firstsection{Introduction}

\maketitle

A new generation of interactive visualization authoring systems has emerged which share a common goal: enabling expressive authoring of visualizations without burdening authors with low-level concerns. They  contrast with prior visualization systems that leveraged chart templates, which provide a predefined palette of chart types with only a handful of customization options, powerful libraries that require programming expertise~\cite{bostock2011d3}, or declarative grammars that require textual specification~\cite{satyanarayan2016reactive, Satyanarayan2017VegaLite}. Yet, despite this shared goal, these systems make notably different design decisions as they trade off two design dimensions: \emph{expressivity}, or what visualizations can be created; and \emph{learnability}, or how difficult is it to author them? Given these differences, traditional evaluative methods are difficult to conduct. For instance, comparative studies are, at best, limited to a small subset of expansive capabilities and workflow permutations that these systems afford, and it is unclear how to fairly quantify and equate different interaction models.

In response, we present \emph{critical reflections} of three systems from this new generation\,---\,Lyra~\cite{satyanarayan2014lyra}, Data Illustrator~\cite{liu2018data}, and Charticulator~\cite{ren2019charticulator}. Our goal is to extract meaningful lessons about their design choices and, as their creators, we are uniquely poised to conduct the necessary in-depth comparisons of their underlying frameworks, system architectures, and interface design. Critical reflection does not fit the typical pattern for evaluation methods as it requires a completed system and the existence of other related systems; these are impossible when publishing a novel system. However, when the potential situation arises, collaborative critical reflections can reveal definitive strengths, weaknesses, and design rationales between the various systems.

Our critical reflections begin by comparing system components. We find areas of broad overlap (e.g., in the set of marks available for use), radically different approaches (e.g., the degree to which scales are exposed, and how complex layouts are interactively specified), and instances where systems have incorporated each other's approach (e.g., Charticulator offers two distinct data binding mechanisms drawn from Lyra and Data Illustrator, respectively). Through these reflections, we distill a set of assumptions that underlie and cut-across all three systems. Reflecting on these assumptions, we realize they were necessary to scope our initial research effort, and discuss how relaxing them suggests exciting opportunities for future work. 

We hope this paper can be an example of a different type of visualization contribution~\cite{lee2019contribution}: a constructive set of lessons learned through critical reflection of the development and public deployment of a set of different but related complex interactive systems.

\section{Related Work}

In this section, we first present an overview of visualization design and authoring and summarize two types of approaches---programming and interactive systems---in visualization authoring.
We also briefly describe the prior theoretical concepts that guided our critical reflections.

\subsection{Approaches to Visualization Design}

\emph{Visualization design} is the general process of visualizing data.
For the purpose of data analysis, it includes understanding the domain problem, the dataset, the task, the visual representation, as well as the algorithms and computer programs for realizing the design~\cite{munzner2014visualization}.
For communicative visualization, the design process includes generating findings from data, creating the visual representations, as well as annotating the visualizations and articulating them with supplemental material to clearly convey the findings.
Visualization design often includes the exploration of design possibilities in an iterative process;  designers make decisions on visual encodings, data content, and the overall composition~\cite{Bigelow2014DesignWithData}.

We consider the narrower activity of \emph{visualization authoring}, 
where the author already has a desired visual representation in mind, and has a dataset in the appropriate format.
However, here, we discuss considerations for visualization authoring, as understanding this process was crucial to the development of our three authoring systems.


Data sketching offers an unrestricted approach to represent data by physically drawing visualizations that span in fidelity: from throw-away sketches to finished compositions~\cite{Walny2015DataSketching,robert2016FiveDesignSheet}.
Related to sketching, vector graphic design applications support rapid, computer-supported design through scene graph representation, composition tools, and transformation actions.
Bigelow et al. analyzed the visualization design approach of graphic designers in~\cite{Bigelow2014DesignWithData}, highlighting flexibility over generative capabilities.
Mendez et al. reconcile this tension by proposing a ``bottom-up" approach for visualization design, as opposed to the ``top-down" approach seen in other visualization design systems~\cite{mendez2017bottomup,mendez2018agency}.

Recently, Elijah Meeks proposed a new way to categorize design approaches based on the underlying convictions and motivations of the data visualization community.
Meeks defines these modes as the ``Three Waves'' of data visualization~\cite{Meeks2018}.
The 1\textsuperscript{st} wave focused on Tufte-esque clarity, simplicity, and thoughtful composition; the 2\textsuperscript{nd} wave focuses on encoding of information based on Wilkinson's Grammar of Graphics~\cite{wilkinson1999grammar} in pursuit of mass-production of visualizations; and the 3\textsuperscript{rd} and current wave is defined by convergence.
Meeks argues that a shift has occurred: moving away from designing individual charts and instead converging on the construction, evaluation and delivery of information products as a whole, including but not defined by charts.
In other words, this third wave coincides with a break from chasing the expanding horizon of what visualizations are capable of, and learning from other visualization design approaches to produce better products. 
Giorgia Lupi echos this sentiment in her ``Data Humanism'' manifesto~\cite{lupi2017data}: \emph{``[the first] wave was ridden by many in a superficial way, as a linguistic shortcut to compensate for the natural vertigo caused by the immeasurable nature of Big Data.''}
In the current wave, Lupi calls for more meaningful and thoughtful visualization with a focus on the complexity, context, and imperfection of data.
In this paper, we take heed of this new wave of data visualization: our reflections take a critical look at the design trade-offs made in pursuit of extending expressive capabilities.


\subsection{Computer-Assisted Visualization Authoring}
We distinguish between programming-based and interactive approaches to computer-assisted visualization authoring.

\subsubsection{Imperative and Declarative Programming}

A wide variety of frameworks, libraries, and declarative languages for creating visualizations with textual programming or specification have been created.
Further, multiple widely adopted visualization programming toolkits and libraries such as Processing~\cite{Reas2006Processing}, D3~\cite{bostock2011d3}, and Vega~\cite{satyanarayan2016reactive} exist. While such toolkits and libraries achieve the ultimate expressivity, full mastery requires considerable programming expertise. 
\emph{The Grammar of Graphics}~\cite{wilkinson1999grammar} forms the basis of many declarative visualization grammars, including ggplot~\cite{Wickham2006ggplot} and Vega-Lite~\cite{Satyanarayan2017VegaLite}. Visualization grammars allows authors to use concise and uniform languages to represent a wide variety of visualizations.
While easier to write than imperative approaches, declarative grammars still pose a challenge as authors must use text to describe interactive visualizations.


\subsubsection{Interactive Systems}

In 2013, Grammel \etal~\cite{Grammel2013Survey} surveyed the landscape of {\it visualization construction user interfaces}\,---\,roughly synonymous to our term interactive visualization authoring systems. 
While many new systems have emerged since this survey's publication~\cite{kirk2019,rost2016}, its distinction between {\it template editors} and {\it shelf construction systems} remains valid.

\emph{Template editors} describe systems like Microsoft Excel, whose expressivity is limited to the set of available chart types. Some systems allow extensive customization by providing a variety of fonts, color schemes, mark types, and graphical styles.
Others such as RAWGraphs~\cite{Mauri2017RAW} and Flourish~\cite{FlourishStudio} provide APIs for developing new templates, though as with the programming approaches surveyed above, this option is out of reach for authors lacking programming expertise.

\emph{Shelf construction} is adopted in commercial systems such as Tableau (formerly Polaris~\cite{Stolte2002Polaris}) and in research prototypes such as Polestar and Voyager~\cite{Wongsuphasawat2016Voyager}.
Instead of using predefined templates, authors map data fields to encoding channels (\eg x, y, color, shape). The system is responsible for generating a valid chart given the author's shelf specification.
However, these systems do not provide control over the underlying chart layout, nor do they allow authors to easily produce compound glyphs comprised of multiple marks.

Finally, another class of interactive systems can be described as {\it visual builders}, and our three systems are members of this class, along with systems such as iVisDesigner~\cite{Ren2014iVisDesigner}, iVolver~\cite{Mendez2016iVolver}, InfoNice~\cite{Wang2018}, Data-Driven Guides~\cite{Kim2017DataDrivenGuides}, DataInk~\cite{Xia2018DataInk}, and VisComposer~\cite{Mei2018VisComposer}.
Though their interfaces tend to be more complicated than those of template- and shelf-based systems, visual builders provide authors with fine control in specifying marks, glyphs, coordinate systems, and layouts.

\subsection{Guiding Design Considerations}

To guide comparison of Lyra, Data Illustrator, and Charticulator, we draw on prior theoretical concepts for evaluating interactive systems. From cognitive science, both Hutchins \etal~\cite{hutchins:direct} and Norman~\cite{norman:doet} define the \emph{gulfs of execution and evaluation} to refer to the difference between a person's intentions and the allowable actions (execution) and the amount of effort that the person must exert to interpret the state of the system and to determine how well the expectations and intentions have been met (evaluation). To bridge these gulfs, one must cover both a \emph{semantic distance}\,---\,the potential mismatch between mental models and the system model in forming one's intention\,---\,and an \emph{articulatory distance}\,---\,the effort required to express that intention~\cite{hutchins:direct}.

Focusing on user interface construction systems, Myers \etal~\cite{myers:past} define the \emph{threshold} and \emph{ceiling} as how difficult it is to learn how to use the system (threshold), and the expressive gamut of what can reasonably be achieved using the system (ceiling). Along similar lines, in the design of the Protovis~\cite{Bostock2009Protovis} and D3 systems~\cite{bostock2011d3}, Bostock \etal posit high-level design considerations of \emph{expressiveness} (``Can I build it?''), \emph{efficiency} (``How long will it take?''), and \emph{accessibility} (``Do I know how?''). We use these same considerations in our comparisons, but use the term \emph{learnability} instead of \emph{accessibility} to avoid confusion. 

We also employ the Cognitive Dimensions of Notation framework~\cite{blackwell2003notational}, which provides a set of heuristics for evaluating the effectiveness of notational systems such as programming languages and visual interfaces. Example cognitive dimensions include \emph{closeness of mapping} (semantic distance of tool representation to domain concepts), \emph{hidden dependencies} (are important links between entities visible?), \emph{hard mental operations} (high demand on cognitive resources), and \emph{abstraction} (types and availability of abstraction mechanisms). This framework has been used to partially evaluate visualization programming approaches such as Protovis~\cite{Bostock2009Protovis} and Reactive Vega~\cite{satyanarayan2014declarative}.

In our subsequent critical reflection, we draw on these concepts to help identify and contrast salient differences in our visualization authoring systems, as well as strengths and weaknesses shared by all three. In particular, we focus on the gamut of visualization designs that can be created with a given system, what domain constructs are exposed for expressing such designs, and the physical interactions and mental operations required for forming, as well as reusing, those expressions.

\section{Background}
\label{sec:scenario}

Our critical reflections focus on the design of three interactive systems for authoring expressive visualizations for communication purposes: Lyra~\cite{satyanarayan2014lyra}, Data Illustrator~\cite{liu2018data}, and Charticulator~\cite{ren2019charticulator}. These systems use a vocabulary that is based on widely used visualization terms (such as marks, scales, axes, and layout) but subtle differences exist in how each system defines these terms, or introduce new ones. Here, we briefly recap the core functionality of the three systems 
by describing how they can each be used to recreate \emph{A Field Guide to Red and Blue America}, a visualization originally published in \emph{The Wall Street Journal}~\cite{wsj:redblueamerica}. We do not expect readers to understand each individual detail, but rather wish to lay the foundation for our subsequent discussion on system components, which sometimes refers back to this example. Video demonstrations of the individual steps and a more comprehensive comparison of terminology are available at \url{https://vis-tools-reflections.github.io}.

The visualization uses bar charts to depict how each state's partisan lean has shifted over time. The bar charts are positioned in a grid structure that mimics the geographic location of their corresponding states. The dataset comprises six columns: the \texttt{State}, \texttt{GridX}- and \texttt{GridY}-coordinates for where this state lies in the grid, \texttt{Year} and \texttt{PVI} which are annual measurements of how the state voted relative to the nation as a whole, and an \texttt{Inclination} field to indicate whether that measurement represents a more Democratic or Republican lean. The original visualization and an excerpt of the data are shown in Figure~\ref{fig:RedBlueAmerica}.

\figureRedBlueAmerica

\bstart{Lyra} We drag a rectangle mark from the toolbar to the canvas. To produce one bar chart per state, we add a ``Group By'' transformation and drag the \texttt{State} field to its property inspector. Doing so introduces a corresponding group mark and nests our rectangle mark within it. We position these groups by dragging the \texttt{GridX} and \texttt{GridY} fields to the width and height drop zones respectively. Doing so sets both the position and dimensions of the groups by producing the necessary scales and axes (we manually remove the latter using the right-hand-side listing). These steps give us a single rectangle for each state, positioned in the grid. To create the bar chart, we drag \texttt{Year} and \texttt{PVI} to the rectangle's width and height drop zones respectively, and remove generated axes. We click to edit the newly generated y-scale, and adjust its range in the configuration panel to depict bars rising or falling. To color the bars, we first add a ``Formula'' transform that calculates an \texttt{Inclination} using a ternary expression (\texttt{PVI > 0 ? 'Rep' : 'Dem'}). Next, we drag this field to the rectangle's fill color and adjust the range of the generated scale to color bars red or blue. Finally, we drag a text mark and drop it over the selected rectangle's top-middle connector. This action produces one text mark instance per rectangle, and anchors the two together. We drag \texttt{State} to the content drop zone to correctly label each chart. In the text mark's configuration panel, we toggle a checkbox to ensure only one label is produced per state, and adjust its appearance (e.g., using boldface and a black font color).

\bstart{Data Illustrator}
We begin by selecting the ``Rectangle'' shape, and drawing a rectangle on the canvas. We click the ``Repeat'' button, select the \texttt{State} field from the drop-down menu, and use the handle to drag out a collection of rectangles, roughly one-per-state. To produce a bar chart, we click into this collection to select a single rectangle, and then click the ``Partition'' button. In the dialog box, we select the \texttt{Year} field to generate one rectangle per-year, per-state. To correctly position these bar charts, we select the collection of rectangles and click the ``Break Grid'' button. Under the ``Composite Position'' panel, we click the data binding icon to map the x and y properties to \texttt{GridX} and \texttt{GridY} respectively. We drag the x- and y-axis handles to produce a larger visualization, and drag the peer counter slider out to display all states in the collection. To color the bars, we select an individual rectangle and bind the \texttt{Inclination} field to its fill color. A color legend is automatically added, and we click each legend item to customize the color palette used. To determine the bar heights, we use the ``Direct Select'' tool and click on the top line segment of a rectangle. In the ``Segment Position'' panel, we bind the y property to \texttt{PVI}, and drag the y-axis handle to ensure bars rise or fall as appropriate. We follow a similar set of steps to label the charts: adding a text mark to the canvas, repeating it by \texttt{State}, and breaking the grid. We bind the x and y positions to \texttt{GridX} and \texttt{GridY}, and choose to reuse the existing scales in the data binding menu. Finally, we bind \texttt{State} to the text content.

\bstart{Charticulator} We begin by dragging a rectangle mark from the toolbar to the glyph editor. This step generates one rectangle per tuple in the dataset. To lay them out as a series of state bar charts, we drag \texttt{GridX} and \texttt{GridY} to the x- and y-axis drop zones of the plot segment (the central region in the chart canvas), select it, and choose the ``Stack X'' sub-layout option from the toolbar. In the configuration panel, we click to toggle the visibility of the x- and y-axes. Dragging \texttt{PVI} to the rectangle's height drop zone in the glyph editor, and \texttt{Inclination} to its fill color property inspector, correctly sizes and colors the rectangles. Clicking the data binding icon reveals a configuration panel to customize the color palette. To label each state, we select the text mark and, in the glyph editor, click the top of the rectangle mark to anchor the two together. In the ``Attributes'' panel, we click the ``Conditioned By'' button, click the \texttt{Year} field in the drop-down menu, and choose the first year. These steps ensure there is only one label generated for each state. Finally, we drag \texttt{State} to the text property inspector, and make adjustments to its position.
\section{Critical Reflections on our System Components}
\label{sec:sys-comp}

We define \emph{critical reflections} as informal discussions by system or toolkit builders to concretely define their collective objectives in support of the user community, and candidly assess the ways each system meets or falls short of these objectives.
In our case, all system builders met weekly for 1 to 2 hour video conference meetings over the course of 3 months.
During these meetings we directly compared our three systems: commenting on our design and implementation, reflecting on practical feedback from the user community, and addressing missed or unexplored research directions.
To structure these conversations, we began by considering the eight evaluative dimensions proposed by Ren et al.~\cite{ren2018reflecting}.
We documented these discussions by recording meeting notes in a shared online folder.
At times, comparing the systems required isolated, preliminary reflection on the individual authoring systems which we documented in the shared folder.
Each team carried out these isolated reflections as ``take-home'' tasks before the next weekly meeting.
These isolated activities provided the due time to exhaustively consider the ways in which each system met or fell short of our defined objectives.

Through this process, we collectively identified \emph{expressivity} and \emph{learnability} as  pervasive dimensions.
Our process also has the benefit of drawing on our collective experience to find relevant consensus and build shared vocabulary.
In this section, we detail the main components the three systems expose and how users create and manipulate them (see Table~\ref{table:components_summary} for the summary).

\begin{table*}[t!]
\caption{Summary of the main components for the three systems. \protect\includegraphics[width=2.5mm]{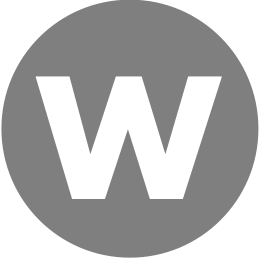} denotes \textit{what} features each system supports and \protect\includegraphics[width=2.5mm]{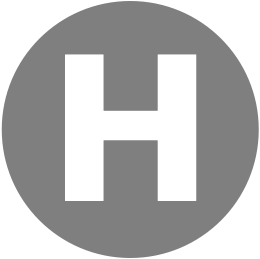} denotes \textit{how} these components are invoked through the user interface. Video demonstrations are available at \url{https://vis-tools-reflections.github.io}.}

\begin{small}
\begin{tabularx}{\textwidth}{ >{\raggedright\arraybackslash}p{.17\textwidth} | >{\raggedright\arraybackslash}p{.24\textwidth} | >{\raggedright\arraybackslash}p{.24\textwidth}| >{\raggedright\arraybackslash}p{.24\textwidth} }

    \toprule
     & \textbf{Lyra} & \textbf{Data Illustrator} & \textbf{Charticulator} \\
    \midrule
    
    \textbf{Mark Instantiation \& Customization} & 
    
    \makecell[tl]{\raisebox{-0.2\totalheight}{\includegraphics[width=2.5mm]{figures/what.png}}
    predefined marks only\\ 
    \raisebox{-0.2\totalheight}{\includegraphics[width=2.5mm]{figures/how.png}}
    drag from toolbar}
    & 
    \makecell[tl]{\raisebox{-0.2\totalheight}{\includegraphics[width=2.5mm]{figures/what.png}}
    custom vector shapes\\ 
    \raisebox{-0.2\totalheight}{\includegraphics[width=2.5mm]{figures/how.png}}
    draw with Pen tool; manipulate path \\points \& segments with Direct Selection tool}
    & 
    \makecell[tl]{\raisebox{-0.2\totalheight}{\includegraphics[width=2.5mm]{figures/what.png}}
    predefined marks only\\ 
    \raisebox{-0.2\totalheight}{\includegraphics[width=2.5mm]{figures/how.png}}
    drag from toolbar; draw in glyph editor}
     \\
    \midrule
    \textbf{Glyph Composition} 
    & 
    \raisebox{-0.2\totalheight}{\includegraphics[width=2.5mm]{figures/how.png}}
    create layers
    & 
    \raisebox{-0.2\totalheight}{\includegraphics[width=2.5mm]{figures/how.png}}
    group marks
    & 
    \raisebox{-0.2\totalheight}{\includegraphics[width=2.5mm]{figures/how.png}}
    compose in glyph editor \\
    \midrule
    \textbf{Path Points \& Path Segments} 
    &
    \raisebox{-0.2\totalheight}{\includegraphics[width=2.5mm]{figures/how.png}}
    map data values to point x, y positions
    & 
    \raisebox{-0.2\totalheight}{\includegraphics[width=2.5mm]{figures/how.png}}
    partition line marks; draw with Pen tool
    & 
    \raisebox{-0.2\totalheight}{\includegraphics[width=2.5mm]{figures/how.png}}
    connect glyphs using the Linking tool
    \\
    \midrule
    \textbf{Links between Glyphs} 
    & 
    \raisebox{-0.2\totalheight}{\includegraphics[width=2.5mm]{figures/how.png}}
    add a force-directed layout transform & 
    not supported 
    & 
    \raisebox{-0.2\totalheight}{\includegraphics[width=2.5mm]{figures/how.png}}
    connect glyphs using the Linking tool \\
    
    \midrule
    \textbf{Data Scoping for Glyphs} 
    & 
    \makecell[tl]{\raisebox{-0.2\totalheight}{\includegraphics[width=2.5mm]{figures/what.png}}
    one or more tuples per glyph\\ 
    \raisebox{-0.2\totalheight}{\includegraphics[width=2.5mm]{figures/how.png}}
    no user action needed for one tuple per \\glyph; ``group by'' for multiple tuples per \\glyph}
    & 
    \makecell[tl]{\raisebox{-0.2\totalheight}{\includegraphics[width=2.5mm]{figures/what.png}}
    one or more tuples per glyph\\ 
    \raisebox{-0.2\totalheight}{\includegraphics[width=2.5mm]{figures/how.png}}
    repeat or partition}
    & 
    \makecell[tl]{\raisebox{-0.2\totalheight}{\includegraphics[width=2.5mm]{figures/what.png}}
    one or more tuples per glyph\\ 
    \raisebox{-0.2\totalheight}{\includegraphics[width=2.5mm]{figures/how.png}}
    no user action needed for one tuple per \\glyph; ``group by'' for multiple tuples per \\glyph}
    \\
    
    \midrule
    \textbf{Mapping Data Values to Visual Properties} 
    & 
    \raisebox{-0.2\totalheight}{\includegraphics[width=2.5mm]{figures/how.png}}
    drag data fields to property inspector or drop zone
    &
    \raisebox{-0.2\totalheight}{\includegraphics[width=2.5mm]{figures/how.png}}
    choose data field from drop-down menu 
    & 
    \raisebox{-0.2\totalheight}{\includegraphics[width=2.5mm]{figures/how.png}}
    perform either method \\
    
    \midrule
    \textbf{Scales} 
    & 
    \makecell[tl]{\raisebox{-0.2\totalheight}{\includegraphics[width=2.5mm]{figures/what.png}}
    all D3/Vega scales\\ 
    \raisebox{-0.2\totalheight}{\includegraphics[width=2.5mm]{figures/how.png}}
    a scale is created when a data binding is \\applied; a scale can be created manually, \\and exists independently from data binding}
    & 
    \makecell[tl]{\raisebox{-0.2\totalheight}{\includegraphics[width=2.5mm]{figures/what.png}}
    scales for categorical, temporal, and \\numerical data\\ 
    \raisebox{-0.2\totalheight}{\includegraphics[width=2.5mm]{figures/how.png}}
    a scale is created when a data binding is \\applied; users choose whether to reuse or \\merge a scale from a previous data binding}
    & 
    \makecell[tl]{\raisebox{-0.2\totalheight}{\includegraphics[width=2.5mm]{figures/what.png}}
    scales for categorical, temporal, and \\numerical data\\ 
    \raisebox{-0.2\totalheight}{\includegraphics[width=2.5mm]{figures/how.png}}
    a scale is created when a data binding is \\applied; by default, reuses a scale from a \\previous data binding}\\
    
    \midrule
    \textbf{Axes \& Legends} 
    & 
    \raisebox{-0.2\totalheight}{\includegraphics[width=2.5mm]{figures/how.png}}
    an axis or legend is created when a data binding is applied; they can be created independently from scales; properties are customized in configuration panel
    & 
    \raisebox{-0.2\totalheight}{\includegraphics[width=2.5mm]{figures/how.png}}
    an axis or a legend is created when a data binding is applied; they can be manipulated through on-canvas interaction, which changes the underlying scale
    & 
    \raisebox{-0.2\totalheight}{\includegraphics[width=2.5mm]{figures/how.png}}
    an axis or a legend is customized through configuration panel; a legend needs to be explicitly added through a button-click \\
    
    \midrule 
    \textbf{Relative Layout} 
    & 
    \raisebox{-0.2\totalheight}{\includegraphics[width=2.5mm]{figures/how.png}}
    drag a target mark to a host mark's anchor 
    & 
    \raisebox{-0.2\totalheight}{\includegraphics[width=2.5mm]{figures/how.png}}
    specify through grouping, distribution, and alignment
    &
    \raisebox{-0.2\totalheight}{\includegraphics[width=2.5mm]{figures/how.png}}
    specify through anchors, guides, handles, margins, and alignment
    \\
    
    \midrule
    \textbf{Layout in a Collection} 
    & 
    \makecell[tl]{\raisebox{-0.2\totalheight}{\includegraphics[width=2.5mm]{figures/what.png}}
    stacking, force-directed, cartographic\\ projections, and pie charts\\
    \raisebox{-0.2\totalheight}{\includegraphics[width=2.5mm]{figures/how.png}}
    add a data transformation via button-click}
    &
    \makecell[tl]{\raisebox{-0.2\totalheight}{\includegraphics[width=2.5mm]{figures/what.png}}
    grid and stacking\\ 
    \raisebox{-0.2\totalheight}{\includegraphics[width=2.5mm]{figures/how.png}}
    apply repeat/ partition actions}
    & 
    \makecell[tl]{\raisebox{-0.2\totalheight}{\includegraphics[width=2.5mm]{figures/what.png}}
    grid, stacking, and circle packing\\ 
    \raisebox{-0.2\totalheight}{\includegraphics[width=2.5mm]{figures/how.png}}
    use scaffolds and sub-layouts to position\\ glyphs} \\
    
    \midrule
    \textbf{Nested Layout} 
    &
    \raisebox{-0.2\totalheight}{\includegraphics[width=2.5mm]{figures/how.png}}
    ``group'' marks are created when dragging data fields to ``group'' drop zones or applying a ``group by'' data transformation
    & 
    \raisebox{-0.2\totalheight}{\includegraphics[width=2.5mm]{figures/how.png}}
    concatenate repeat and partition actions in a flexible order 
    & 
    \raisebox{-0.2\totalheight}{\includegraphics[width=2.5mm]{figures/how.png}}
    map categorical data columns to X \& Y axes and apply sub-layout; embed a Charticulator template as a nested chart \\
    
    \midrule
    \textbf{Coordinate Systems} 
    & 
    \raisebox{-0.2\totalheight}{\includegraphics[width=2.5mm]{figures/what.png}}
    Cartesian 
    & 
    \raisebox{-0.2\totalheight}{\includegraphics[width=2.5mm]{figures/what.png}}
    Cartesian 
    & 
    \makecell[tl]{\raisebox{-0.2\totalheight}{\includegraphics[width=2.5mm]{figures/what.png}}
    Cartesian, polar, and arbitrary curve\\ 
    \raisebox{-0.2\totalheight}{\includegraphics[width=2.5mm]{figures/how.png}}
    drag from toolbar into a plot segment}
     \\
    
    \bottomrule
\end{tabularx}
\end{small}
\vspace{-3mm}
\label{table:components_summary}
\end{table*}

\subsection{Marks}

Instantiating \emph{marks} is a fundamental operation in visualization authoring systems. A mark is a primitive graphical element in a visualization, which includes rectangle, symbol, line, arc, image, and other shapes.

In Lyra, an author may instantiate a mark by dragging the corresponding icon from the marks toolbar to the canvas. Data Illustrator uses a toolbar similar to vector graphics editing applications. An author instantiates a mark by activating the desired tool, and then draw on the canvas by mouse clicks and drags (depending on the system). Charticulator supports both approaches. While the drag-and-drop approach increases directness by requiring just one single action to create a mark, the system will need to provide a reasonable default location and size for the mark. When such choice is not valid, the author may see unexpected results. On the other hand, Data Illustrator's interaction introduces a state of the currently activated tool. When the author loses track of this state, s/he may unexpectedly create an unwanted mark. Charticulator relieves this problem by making the activated state one-off\,---\,after creating the mark, the system deactivates the mark tool.

Authors sometimes need to create a \textit{glyph} consisting of one or more marks. In Data Illustrator, this is done by drawing marks on the canvas and grouping them. Lyra similarly offers a ``group'' mark. Charticulator introduces a special canvas called the \emph{glyph editor}, and marks can be instantiated in either the glyph editor or the chart canvas. Providing a separate glyph editor affords a larger editing region for a glyph, and makes it easier to select and manipulate marks. However, it introduces a deviation from conventional single canvas authoring systems. This design also requires the author to be aware that what is shown in the glyph editor is a prototype, and to predict the effect of an action on instances other than the currently selected one. Separating interactions between a glyph editor and a chart canvas also creates confusion, as the author will need to know the roles of the two canvases, and decide where to select and manipulate an object.


While Lyra and Charticulator rely on predefined mark types such as a rectangle, symbol, or text, Data Illustrator supports a wide spectrum of custom shapes by providing a ``pen'' tool\,---\,a common feature in vector graphics software.
All shapes besides text and image are represented as paths. A path consists of multiple path points and path segments. An author can freely manipulate path points to achieve a desired shape. In addition, the location of path points and path segments can be bound to data. While this might be hard to learn for authors unfamiliar with vector graphics editing software, having such flexibility over mark shapes increases Data Illustrator's expressiveness.


\figureDataBinding

In Lyra and Data Illustrator, marks include not only individual elements, but also polylines or curves that connect multiple data points. Lyra supports line and area marks. The $(x, y)$ position for such marks can be bound to data to create lines or areas that connect through these points. In Data Illustrator, a line can be partitioned by a data column, resulting in a line with multiple path points. The $(x, y)$ coordinates of such path points can be bound to data. In Charticulator, however, glyphs (which may contain multiple marks) are by design individual graphical elements. Each glyph corresponds to one or a group of data tuples. To create visual links \emph{between} glyphs, authors need to use the ``linking'' tool, which draws lines or bands between glyphs. This separation of glyph and link introduces a level of indirectness that impacts learnability, as even basic line charts must use the link tool. However, it increases Charticulator's expressiveness, making it possible to create charts that have links, such as chord diagrams and arc diagrams.

\subsection{Data Binding}
\label{sec:data_binding}

Data binding is a core operation in authoring visualizations that involves (1) generating glyphs based on data, 
and (2) specifying a mapping between data fields and mark properties such as position, color, or size.

\subsubsection{What Data Does a Glyph Represent?}
\label{sec:mark-tuple-mapping}
In Lyra and Charticulator, a glyph represents one data tuple, respectively. Authors thus need to prepare the dataset to fit this assumption. For example, to create a bar chart, each row in the dataset must correspond to one bar. Based on this assumption, the generation of glyphs is automatic in these two systems. Whenever a glyph is updated in the Glyph Editor in Charticulator, the system automatically generates glyphs, one per tuple; similarly, in Lyra, glyph generation by data is not a user operation. The underlying Vega grammar follows a \textit{declarative} approach, and considers explicit glyph generation to be \textit{imperative}. Glyphs are thus automatically generated by Vega when authors specify a mapping between a data variable and a visual property.

Data Illustrator relaxes the requirement of one-to-one glyph-tuple mapping, and allows a glyph to represent one or more tuples through two core operations: repeat and partition. Using the \emph{Field Guide to Red and Blue America} dataset as an example (Section \ref{sec:scenario}), one can repeat or partition a rectangle using any of the categorical data variables. Repeating a rectangle by \texttt{State} will generate 51 rectangles (50 states and D.C.), and each rectangle represents all the tuples sharing the same \texttt{State} value. The tuples represented by each glyph are called the \textit{data scope} of the glyph. The notion of data scope applies to path points and segments too. We can partition a line using \texttt{State}, and each point on the resultant path represent the corresponding tuples as its data scope.

Data Illustrator's approach allows authors to dynamically aggregate data via direct manipulation in a more uniform fashion than the other two systems. For instance, authors can create a bar chart where each bar represents a \texttt{State} and the height encodes the average \texttt{PVI} values. To perform the same operation in Lyra, authors need to instantiate a ``Group By'' data transformation followed by a ``Stats'' calculation via the left-hand side configuration panels. Similarly, in Charticulator, authors need to specify a ``Group By'' attribute on the plot segment, which aggregates the data with user-selectable aggregators. 

This improved expressivity, however, introduces an additional layer of complexity as authors must verify and keep track of the glyphs' data scopes. To address this issue, Lyra and Data Illustrator dynamically update the data table panel to reflect the data scope of selected marks.

These differences in glyph-tuple mapping have significant implications on authoring nested layouts, which we discuss in Section~\ref{sec:nested_layout}.


\subsubsection{Mapping Data Values to Visual Properties}



Lyra offers two ways for constructing a data binding: data fields can be dragged to a property's inspector or to a property \emph{drop zone}, a shaded region that overlays the visualization canvas. Data Illustrator eschews drag-and-drop interaction in favor of button-clicks. Properties that can be data-driven display a \emph{binding icon} alongside their inspector; clicking the icon reveals a drop-down menu of data fields that the author can choose from. Charticulator adopts both approaches\,---\,fields can be dragged to drop zones 
or can be chosen from a drop-down menu revealed by clicking a property's binding icon. Figure~\ref{fig:DataBinding} illustrates these different approaches.

The differences in these data binding interactions are grounded in learnability. Lyra's design sought to be familiar to Tableau users, while also increasing their sense of directness. Thus, dropping fields to property inspectors recalls Tableau's ``shelves'' metaphor~\cite{Stolte2002Polaris} while adding additional drop zones to the visualization canvas. Overlaying drop zones directly over the properties they correspond to narrows the gulf of execution~\cite{norman:doet}. A first-use study confirmed these effects as participants described drop zones as \emph{``natural''} and \emph{``intuitive''} and, when compared to Tableau's shelves, made them feel more in control. However, participants also identified two weaknesses in this interaction model: (1) drop zones provide a small active region that can be difficult to hit consistently; and (2) drop zones rely on a having a mark selected, which determines how drop zones are depicted, and can be difficult to keep track of after a data binding occurs.

Data Illustrator was intentionally designed to address these shortcomings, and identified two additional concerns. Tableau and Lyra ask authors to perform dragging operations over potentially long distances, which yields a poor experience in terms of both efficiency and accessibility (e.g., for users with motor impairments). Moreover, neither system makes it clear if particular data bindings would result in inexpressive or ineffective outcomes~\cite{mackinlay1986automating} (e.g., binding a quantitative field to an identity channel such as shape~\cite{munzner2014visualization}). Data Illustrator's binding icon, in comparison, provides a single interface element that is consistently displayed regardless of the particular property being targeted. Clicking the icon, and selecting a field from the drop-down menu, is more efficient than performing a drag operation, and available fields are filtered to ensure only valid data bindings can be constructed. These gains, however, are offset by a loss of directness\,---\,data binding is the only core operation in Data Illustrator that cannot be specified by manipulating the visualization itself.

Charticulator makes the fewest compromises by adopting both the binding icon and drop zone model, and refining the latter in a few crucial ways. When one begins to drag a field, only valid drop zones are shown (and corresponding property inspectors highlighted) to ensure that improper data bindings cannot be constructed. And, the drop zones are visualized in the glyph editor, providing a single consistent place in the interface for interacting with them rather than Lyra's dependence on the currently selected mark instance. As a result, Charticulator is able to maintain Data Illustrator's efficiency and accessibility advantages, without sacrificing Lyra's directness. Echoing Lyra's evaluation, participants in Charticulator's usability study rated drag-and-drop interaction as one of the aspects they liked most about the system.

However, there are still opportunities for further improvement. Charticulator's drop zones, like Lyra's, provide a relatively small active region and thus require the author to perform a fairly precise interaction. This design can yield a frustrating experience when a author drops a field close to, but not directly over, a drop zone. Inspired by the bubble cursor~\cite{grossman2005bubble}, an in-development version of Lyra accelerates drop zone acquisition by computing an invisible Voronoi tesselation over valid drop regions. Thus, the drop zone nearest to the mouse cursor is automatically chosen and the author no longer needs to drop directly over a drop zone to successfully establish a data binding.

In all three systems, the outcome of establishing a data binding is immediately reflected on the visualization canvas. This behavior tightens the feedback loop, enabling a more iterative authoring process and reducing the gulf of evaluation~\cite{norman:doet}. Lyra and Data Illustrator, however, identify a further need to bridge this gulf. They note that interactive data binding results in an accretive authoring process that displays intermediate visualization states to the author. For instance, in Lyra, binding a rectangle's height before its x-position or width produces a set of overlapping mark instances. In Data Illustrator, a mark's \emph{data scope} (i.e., the set of tuples it is bound to) may change over time. Both systems scaffold this experience by exposing the backing data in a persistent tabular interface (Charticulator's data table is a modal display). Data Illustrator goes a step further by only displaying records in the table that correspond to the selected mark's data scope.

Although our discussion here has centered on how these data binding interactions are designed to support learnability, there are important expressivity concerns as well. For instance, Lyra's two data binding mechanisms carry different semantics. When a data field is dragged to a drop zone, the system automatically infers the necessary scale functions and adds appropriate axes or legends to the visualization. This inference does not occur if the field is dropped over a property inspector, allowing for more fine-grained design choices (e.g., if the author wishes to use a scale function they have manually constructed, as described in~\ref{sec:scales}) or for authors to bypass scale functions and have visual properties set to data values directly (e.g., when colors carry semantic resonance, as is typical with public transit data, or if data values are produced by algorithmic layouts, see~\ref{sec:layout}). However, Lyra does not provide clear affordances communicating these differences in its interface. When an author releases a field over a property inspector, they may notice that the corresponding interface element for a scale function does not appear, but this difference is subtle and requires them to have a sufficient level of expertise to understand the purpose of scales. One could refine this design for learnability, without sacrificing the expressivity gains, by making it opt-in: by default, dropping fields on property inspectors could still trigger scale inference, but a dialog box could allow more advanced authors to bypass it for future interactions. 

Data Illustrator's data binding also carries subtly different semantics compared to the other two systems. If a data binding is removed, marks do not revert back to their previous, unbound appearance (as they do in Charticulator and Lyra). Instead, they return to being standard vector shapes that can be manipulated via drawing interactions like dragging to resize, rotate, or move. This strategy, called a \emph{lazy} data binding, enables ``fuzzy'' layouts that approximate the original data values and supports an approach that researchers have found occurs commonly in practice, as designers seek to maintain a flexible and rich design process~\cite{Bigelow2014DesignWithData}. While powerful, it is also important to note that this technique is ripe for misuse, making it easier for people to author visualizations that look roughly accurate but with subtle changes introduced. Although it would be impossible for a system to entirely prevent such misleading visualization from being created, there are a number of ways that these outcomes could be mitigated. For instance, the system could warn authors and ask them to confirm their desire to remove a data binding. Marks that are subsequently manipulated could be identified with a persistent warning in their configuration panel, with an option to revert any changes. Or, the output visualization could allow readers to re-establish data bindings and compare the differences for themselves (e.g., via an overlay display or through animation). 
\subsection{Scales, Axes, and Legends}
\label{sec:scales}

Our systems all use three constructs to operationalize data bindings: (1) scales, functions that map the data domain to a range of visual values; (2) axes, visualizations of spatial scales; and, (3) legends, visualizations of scales of non-spatial properties such as color, shape, or size. However, how visible these abstractions are to authors, and how the constituent properties are manipulated, vary significant as the three systems make different tradeoffs between expressivity and learnability. 

\subsubsection{Scale Visibility}

When a data binding interaction is performed, all three systems automatically construct any necessary scale functions. However, they lie along a spectrum with regards to the degree these scales are exposed to the author. At one end lies Lyra, which provides scales as first-class primitives: authors are able to manually construct a scale independent of any data binding interaction. At the other end sits Charticulator, which does not distinguish a scale from its axis or legend\,---\,controls to modify the domain or range are shown on axis or legend configuration panels. And, Data Illustrator lies in-between. When an author clicks the bind icon they are prompted to create a new scale, or reuse an existing scale if the field has been previously used. Similarly, via the same interface, authors can elect to instead merge scales, producing a single scale with a domain unioned across several fields.

The level of visibility has clear implications for expressivity. For example, in Data Illustrator, merging the scale functions enables authoring Gantt Charts where task start and end times are recorded as separate fields. Authors using Lyra have complete control over scale functions. Thus, they can create a scale with a domain that spans multiple distinct data fields, subsequently use it with a field outside of its domain to determine a data binding, and reuse or merge scales in a more fine-grained fashion than in Data Illustrator. This flexibility can be important even for creating simple visualizations. For instance, consider a scatterplot in which the author wants to ensure Euclidean distances are accurately plotted, or an asymmetric adjacency matrix. Both these examples require using a single scale function for both the x and y dimensions, with a domain unioned over two data fields.

However, this expressivity comes with a non-trivial complexity cost. Once an author establishes a data binding, Data Illustrator's interface does not make clear which scale function is being used\,---\,is it a new scale, is it an existing scale, or have scales been merged? Though HCI theory cites maintaining high visibility into system components as an important dimension for reducing the gulf of evaluation~\cite{hutchins:direct, blackwell2003notational}, on-going feedback for Data Illustrator and Lyra indicate that authors struggle to understand the role that scales play. In Lyra, this issue is compounded by the additional user interface elements that come with first-class scales: a separate panel that lists all available scale functions; and, a corresponding interface element for each scale that can be dragged and dropped, and appears alongside data fields as part of a data binding. Ironically, behavior that was designed to reduce interface clutter\,---\,scales, axes, and legends that were automatically created are also automatically removed when they no longer participate in a data binding\,---\,has had the knock-on effect of increasing churn in the user interface. How best to address this complexity, without losing the expressivity gains, remains unclear. An ``advanced'' mode is unappealing, as it introduces additional discovery costs, and would turn this complexity into a cliff rather than smoothing it out. We explore an alternate strategy Lyra might take in the subsequent subsection. 

\subsubsection{Manipulating Axes \& Legends}

Data Illustrator and Charticulator couple a scale to its axis or legend representation. Modifications of axis or legend properties (e.g., specifying an alternate color palette) map to transformations of the underlying scale function. Lyra, on the other hand, provides axes and legends as first-class primitives that can be modified independently from the corresponding scale. Lyra's approach yields expressive gains as well as a less viscous user experience~\cite{blackwell2003notational}\,---\,for instance, axes and legends can be created to visualize scales that do not participate in a data binding, and alternate scales can be chosen without losing any axis or legend customizations. Nevertheless, to address the complexity of scale visibility, Lyra may consider adopting a hybrid approach: scales, axes, and legends remain first-class primitives that can be manually constructed and independently customized, but are coupled if they are created automatically during a data bind.

As Figure~\ref{fig:ScalesAxesLegends} shows, authors customize axes and legends via configuration panels in Lyra and Charticulator while Data Illustrator opts for a more direct interaction model: authors can click and drag to reposition axes and legends; handles overlay axes on hover and can be dragged to shorten or lengthen an axis; and clicking individual entries in a color legend launches a color picker. Data Illustrator's interaction model significantly reduces the articulatory distance of modifying scale functions but also introduces concerns of hidden dependencies~\cite{blackwell2003notational}. As axes and legends provide the sole mechanism for reifying scale functions, if they can be freely repositioned on the canvas, it can be easy to lose track of which marks they correspond to. More problematically, support direct manipulation modification has slowed down progress on expanding expressivity. Although Data Illustrator's underlying framework is capable of expressing axis and legend customizations (e.g., to grid lines, ticks, labels, etc.), how to expose this functionality via direct manipulation remains ongoing work. Lyra's configuration panels, though they afford less of a sense of directness, provide an extensible interface component that surfaces these fine-grained properties in a consistent fashion.  


Charticulator makes an important exception to its treatment of axes and legends: legends, unlike axes, are not automatically added to the visualization. Instead, authors must manually add a legend using a predefined legend element or by creating a new glyph in a separate plot element. Although authors could choose to manually create custom legends in Lyra or Data Illustrator, Charticulator makes this choice more explicit. This approach is designed to recognize that while axes have a mostly uniform appearance (a main horizontal or vertical line, with individual tick lines, labels, and a title), legends have a much more expressive design space.\footnote{A comparison of the axis and legend design spaces, using the Vega visualization grammar~\cite{satyanarayan2016reactive} is available at \url{https://observablehq.com/@vega/a-guide-to-guides-axes-legends-in-vega}.} For instance, a designer may choose to directly label a multi-series line chart rather than use a separate legend. By not automatically adding a legend, Charticulator seeks to reduce an author's propensity for design fixation~\cite{jansson1991design}, but it incurs a non-trivial learnability cost. The button to add a pre-defined element is buried within a configuration panel (see Fig.~\ref{fig:ScalesAxesLegends}), and creating a high-fidelity legend can be as complex as authoring the original visualization. Future versions might lower this threshold~\cite{myers:past} through new abstractions for generating legend entries (e.g., akin to Data Illustrator's repeat).


\figureScalesAxesLegends

\subsection{Layout}
\label{sec:layout}

To enable expressive layout, the systems expose a variety of methods for manual and relative positioning along a coordinate system, as well as collective placement for nested and small multiples displays.


\subsubsection{Relative Layout}
Both Lyra and Charticulator use an \textit{anchor} for relative positioning between visual objects. In Lyra, once authors establish a connection by dragging a target mark onto a host mark's anchor, the target mark’s position is automatically determined by the host's properties. Similarly, Charticulator uses anchors and handles to specify the layout relationship between two objects in both the glyph level (i.e., between marks) and in the chart level (i.e., between plot segments and one-off marks). Moreover, Charticulator's marks have margin and alignment properties that can be used for similar means. For example, a text mark representing a data value for each bar can be placed at the bottom of the bar (inside or outside) or at the top of the bar (with the same distance from the bar). Snapping one mark to another results in a snapping constraint, which remains in effect unless the author proactively unsnaps the mark. These behaviors are not found in vector graphics environments. Despite a learnability cost, these constraints specify more reusable designs.

\subsubsection{Layout in a Collection}
After generating shapes from data (Section \ref{sec:mark-tuple-mapping}), we get a collection of marks/glyphs. In Lyra, these glyphs are placed at the same position, overlapping each other, since they are duplicates of the original glyph prototype. Doing so, however, would not reveal that multiple glyphs have been created and this could confuse people. To address this problem, Charticulator introduces a scaffold to position the glyphs in simple (horizontal or vertical) stacking, grid, and circle packing layouts. Data Illustrator adopts a similar approach. Marks generated by the repeat operation are placed in a grid layout by default, whereas marks generated by the partition operation are stacked by default. The interface allows one to adjust the horizontal and vertical gaps in a grid layout by directly manipulating padding handles. These design choices allow automatic positioning of glyphs in a collection without any specification of mappings between data variables and spatial coordinates. Once an author binds a data variable to the x- and y- positions of marks, an axis is generated and the glyphs are placed based on data values.

\subsubsection{Nested Layout}
\label{sec:nested_layout}

Nested visualizations such as grouped bar charts and small multiples are common. As mentioned in Section \ref{sec:mark-tuple-mapping}, 
the differences between the systems in terms of mark generation and data mapping have ramifications on the creation of nested structures. In Charticulator, once a rectangle is added in the glyph editor, the system automatically generates all the marks, one mark per tuple. The main task is thus to lay out these marks. The basic layout in the \emph{Field Guide to Red and Blue America} example (Fig.~\ref{fig:RedBlueAmerica}) is specified in three steps in Charticulator: bind \texttt{GridX} to the plot segment's x-axis, bind \texttt{GridY} to its y-axis, and apply ``Stack-X'' sub-layout (Fig.~\ref{fig:RedBlueAmericaSteps}). When binding a variable to an axis, Charticulator makes automated decisions: if the variable is categorical, and if multiple marks share the same value, Charticulator will group these marks and apply a default grid sub-layout to arrange them; if the variable is numerical, no grouping will be applied, and the marks will be at the same position, on top of each other. This automated decision requires \texttt{GridX} and \texttt{GridY} to be formatted as strings (e.g., ``I5'') instead of numbers (e.g., 5). Authors unaware of this logic may have difficulties in understanding system behavior. 

Nested layout in Data Illustrator is achieved by combining repeat and partition operations. Figure \ref{fig:RedBlueAmericaSteps} shows one workflow: partition a rectangle by \texttt{Year}, which results in a collection of rectangles; then repeat the collection by \texttt{State}. Alternatively, we can repeat a rectangle by \texttt{Year}, and then repeat the resultant collection by \texttt{State}. The combination of repeat and partition operations is flexible, and requires authors to have a good understanding of how these two operations work.

In Lyra, nested layouts are achieved via ``group'' marks that are automatically instantiated when authors drag data fields to the ``Grouped Horizontally'' or ``Grouped Vertically'' drop zones, or instantiate a ``Group By'' data transformation. These marks, akin to Charticulator's plot segments,  serve as containers for axes, legends, and graphical marks. Thus, to recreate our example, we first add a ``Group By'' data transform and drop \texttt{State} into its property inspector. We then drag \texttt{GridX} and \texttt{GridY} to the resultant group mark's ``width'' and height drop zones, respectively, which both positions and sizes the groups. Like Charticulator, Lyra expects \texttt{GridX} and \texttt{GridY} to be strings to infer a categorical scale for group positioning. Lyra's approach is the weakest of the three systems, as it introduces a set of hidden dependencies that present a highly viscous experience~\cite{blackwell2003notational}. A priori it is not at all clear that group marks exist, and if one wishes to switch between horizontal and vertical layouts, they must remove and recreate the grouping.



%
These different ways of representing and constructing nested layout have implications on component selection. As Lyra maps interactions to declarative statements in the Vega visualization grammar, which does not formally represent graphical components such as collections, there are fewer types of selectable components. In Data Illustrator, the selection mechanism is hierarchical: Clicking once selects the top level collection, double clicking opens up the collection so that the individual marks can be selected. This design closely follows the selection model in Adobe XD. Selecting a mark can be tedious, however, if authors intend to select a mark embedded in multi-level nested collections, multiple double clicks are needed to open up the collections hierarchically. To avoid this problem and reduce visual clutter, Charticulator provides two separate editing canvases: a glyph editor, and a chart canvas. The selection of marks and anchor points can take place in either the glyph editor or the chart canvas. Selecting a glyph or a plot segment solely happens in the chart canvas. 





For more complicated visualizations such as small multiples with multiple levels of nesting, Data Illustrator keeps a consistent user interface, allowing people to apply the repeat operation multiple times. Similarly, Lyra's group marks can be nested arbitrarily deep, by instantiating additional ``Group By'' transforms, or using the corresponding drop zones. Charticulator, on the other hand, extends the notion of a glyph to include a ``nested chart.'' To create a nested chart, one can either import a pre-exported Charticulator template, or use the ``nested chart editor,'' which is essentially a popped-up Charticulator user interface with the nested chart and its corresponding portion of the dataset. Authors can then generate glyphs with ``nested charts'' by data. There are some limitations of this approach, however. Each small multiple instance has its own constraint solver, thus it's not possible to add constraints across the instances. In addition, scale and axis parameters are currently shared across instances of small multiples, but they are inferred from a single small multiple instance. The author has to manually unify scales/axes across all instances.

\figureRedBlueAmericaSteps

\subsubsection{Coordinate Systems}
The Cartesian coordinate system is a fundamental aspect of chart layouts. As such, Lyra, Data Illustrator, and Charticulator all support the Cartesian coordinate system. However, each takes a different approach to address additional coordinate systems or more advanced layouts including algorithmic ones.

Lyra's grammatical primitives only offer compositional expressivity for the basic Cartesian coordinate system and its ``reactive geometry''~\cite{satyanarayan2016reactive} makes it possible to anchor marks together for additional layouts (e.g., stacking, pie charts). Lyra supports more advanced layouts (e.g., treemaps) using modules that can be invoked from button presses/property inspectors, often requiring a specific data format (e.g., geographic, hierarchical, network data); people need to choose from a typology of data transformations. 
In other words, Lyra's different specification options result in an inconsistent interaction model.

In contrast, Data Illustrator's repeat and partition operators offer a consistent mechanism for basic and advanced layouts, but important implementation and conceptual limitations remain. Support for polar coordinates is conceptually well-defined though it has yet to be implemented. For instance, partitioning a circle would produce a pie chart. 
However, how these operators extend to support geographic, hierarchical, or network data is an open question. 

Charticulator's system of constraints offers the most uniform underlying model for layout, enabling coordinate systems such as polar and arbitrary curve coordinates, in addition to Cartesian. Advanced layouts can be implemented as a module and incorporated into Charticulator via an additional panel (as Lyra did). However, because this is not aligned with one of the core design goals (i.e., enabling people to specify a novel layout using a set of partial constraints), we acknowledge these additional algorithmic layouts as a known limitation.

\section{Critical Reflections on our Shared Assumptions}
\label{sec:assumptions}

The previous section explored the ways in which our three systems differ with respect to the tradeoffs between expressivity and learnability.
To complement our reflection on differences and tradeoffs, we now reflect upon our systems' \textit{shared} assumptions 
about the authors who use our systems, the data that authors load into our systems, the tasks that they perform, the system requirements, and the content that authors export from them.
We now have greater clarity regarding these assumptions, assumptions that were not explicitly defined when we set the scope of our research and developed these systems, when our priorities were primarily expressivity and novelty.
We acknowledge that these priorities likely hindered the adoption of these deployed systems to some extent.
Reflecting on these simplifying assumptions provides opportunities for further refinement to drive adoption as well as directions for future research.



\subsection{The Author: Literacy \& Skill Transfer}

A shared assumption underlying our systems is an author's desire to visualize data without programming while still exhibiting a level of comfort with computational thinking; otherwise they would have elected to manually illustrate their charts (e.g., using pen \& paper or a vector graphics application).
We further assume some level of data literacy, as authors would need to understand the structure and type of data that can be loaded into the systems.

Another shared assumption pertains to familiarity with other systems. 
Given Data Illustrator and Charticulator's ties to Adobe and Microsoft, respectively, they may attract authors who are already familiar with other systems produced by these organizations.
For instance, experience using Adobe Illustrator may help people as they learn to use Data Illustrator, while a familiarity with configuring custom visuals for Microsoft Power BI may contribute to the process of learning to use Charticulator.
However, skill transfer may occur from many other sources, and thus attaining a better understanding of skill transfer for learning our systems is an important direction for future research.

\subsection{The Data: Cleaned \& Pre-Processed}
With regards to the data that people load into our systems, our common assumption is that the data is an appropriately formatted CSV, TSV, or JSON file. Based on the example charts showcased in the systems' respective papers and online galleries, the size of the datasets in these files is also small: a handful or columns and dozens of rows, or a few hundred rows at most. There are no missing or erroneous values, and the data has already been appropriately cleaned, filtered, and aggregated using some other set of data preparation and analysis tools. Finally, the data is static, meaning that the structure or values of the data will not change at some later point in the authoring process.

All three systems exhibit an issue that we call \emph{schematic congruency}: authoring a visualization may require the backing dataset to be structured or formatted in a particular way that may not be clear to authors a priori. In particular, all three systems expect datasets to be structured in a long (often referred to as ``tidy,'' and as opposed to wide) format. Data Illustrator and Charticulator do not provide facilities to transform the backing data beyond sorting and calculating summary statistics such as the mean or median. Lyra, on the other hand, provides a palette of data transformations including operators to filter and aggregate the dataset, as well as derive new calculated fields.\footnote{Vega's fold operator, used to transform a wide dataset into a long format, was added after Lyra's initial development; exposing it within Lyra would be a straightforward implementation effort.} An additional set of layout transformations support ingesting non-tabular data (e.g., networks or geographies) but they impose their own set of structuring and formatting concerns that are invisible to the author. For example, the nodes and links of a network must be imported as two separate datasets, with each link described by properties named \texttt{source} and \texttt{target} exactly. 

Although supporting a rich set of data wrangling capabilities~\cite{kandel2011wrangler} is out of scope, how these systems should narrow schematic congruency remains an open question. As Lyra illustrates, it is not sufficient to simply extend the underlying visualization models to provide data transformation capabilities. Such an approach presents a non-trivial gulf of execution~\cite{norman:doet, hutchins:direct} by expecting authors to manually define data transformations. Instead, these systems must develop higher-level scaffolding that automatically infers or suggests appropriate transformations when necessary, analogous to their existing mechanisms for data binding which automatically infer definitions for scales and guides. 

\subsection{The Task: Authoring, not Designing}

Another fundamental assumption is that people want to \textit{author} a chart, not \textit{design} one.
In other words, we assume that users do not wish to explore the visualization design space using our systems but rather come with a specific chart design in mind.
Perhaps their design is something they sketched on paper, or maybe they want to emulate a design they saw elsewhere (e.g., in our systems' associated galleries).
This assumption further implies that the author is already acquainted with their dataset\,---\,we imagine the author to have already performed some exploratory data analysis and preparation prior to using the systems.

It is unclear how these assumptions hold up in practice.
For authors without a specific design in mind, or for those with little or no understanding of their dataset, our systems are of little help as they begin with a blank canvas, a \textit{tabula rasa}.
Moreover, these assumptions yield largely linear interaction flows.
While all systems present several different entry points to author a given design, once one starts down a particular path it can be very difficult to make a change and often involves starting from scratch (a highly viscous user experience~\cite{blackwell2003notational}).
This process stands in stark contrast to how design practice prioritizes a broad exploration of the design space (e.g., through alternating flare and focus phases~\cite{buxton2010sketching} or parallel prototyping~\cite{dow2010parallel}).



Though all three systems have been released, with deployed versions that can be readily used, their adoption has been hindered by poor support for organic practice. 
Thus, relaxing these assumptions may facilitate a more natural design process. For instance, how might our systems allow authors to explore and combine divergent design ideas? How do we adapt the fork/merge workflow, popularized by version control systems like Git, for interactive visualizations?
Or how might our systems incorporate visualization design recommendation~\cite{wongsuphasawat2016compassql}?

\subsection{Export, Reuse, \& Interoperability}

Visualizations appear across a range of media (e.g., presentations, websites, or print articles).
Given this range, authors often add refinements to their visualizations (e.g., adding annotations or visual embellishments) using another application~\cite{Bigelow2014DesignWithData}. Our systems remain agnostic about downstream uses, offering mechanisms to export the visualization being constructed. The most common format is Scalable Vector Graphics (SVG)\footnote{Lyra and Charticulator also offer the ability to export the visualization as a rasterized image (i.e., PNG, JPEG), to simplify embedding the visualization.} to ensure high-fidelity display across a range of screen resolutions, and to support further customization in graphics applications such as Adobe Illustrator~\cite{Bigelow2017Hanpuku}.

However, exporting the visualization as an image is a lossy operation. The link to the backing dataset is broken, and visualization semantics (e.g., whether a vector line denotes a mark or axis tick) is lost. With only an SVG representation, it is difficult to recreate the visualization with different input data, or perform any higher-level reasoning about it (e.g., re-targeting it to different form factors~\cite{Sam2018}). Currently, Data Illustrator does not support such use cases although one can anticipate how its framework could be extended to do so. We instead focus our discussion on Lyra and Charticulator.

Lyra and Charticulator visualizations can be exported as reusable templates\,---\,as a Vega specification~\cite{satyanarayan2016reactive} expressed using JavaScript Object Notation (JSON), or as a custom JSON-template or Power BI custom visual~\cite{PowerBI}, respectively. With these templates, authors can generate new instances of the visualization with different input data, or render it in different-sized canvases and have elements adapt appropriately (e.g., resized scale and axis extents). Charticulator templates can also be imported as a (nested) chart component for a small multiples display.
However, these templates remain hard-coded to the original dataset schema. Any new data must be structured in an identical format (i.e., columns of the same types as in the original dataset). 
\section{Discussion: Reflecting on Reflections}


\bpstart{Critical Reflections are a Viable Evaluation Method}
In this paper, we contribute \textit{critical reflections}, an alternative method of evaluating complex interactive systems motivated by the difficulty of using traditional comparative studies~\cite{ren2018reflecting}. 
Critical reflections are modeled after design critiques\,---\,a popular process in the design community for analyzing how effectively design choices meet goals~\cite{Connor2015DiscussingDesign}.
Unlike critiques, which occur throughout an iterative design process, being retrospective has allowed us to understand the different trade offs that can be made in pursuit of common goals.
We believe this paper demonstrates that critical reflections are a viable and informative evaluation method, providing a constructive set of ``lessons learned'' to inform future research in visualization authoring systems.
Moreover, we imagine that the critical reflection process would be equally applicable to a single system and may yield more valuable insights than reporting trivial usability issues or conducting contrived comparative user studies.
Reflecting on a single system would involve not merely articulating the rationale for the design choices made, but discussing what alternate prototypes were considered, what their relative strengths and weaknesses were, which aspects were incorporated into the final version, or why they were ultimately discarded.


\bpstart{The Devil is in the Details}
We note that, even for the creators of one of the three systems, it would have been infeasible to acquire these insights by simply reading the individual papers.
As there were significant inconsistencies between the abstractions each system exposed, all stakeholders needed to actively participate in extracting and mapping intricate \emph{low-level details}\,---\,the heart of the critical reflections process. 
Detailed discussions of design rationales helped us understand the full extent of the design space and revealed points that were often new to one or more of us.
Distilling this information into a series of high-level takeaways would have defeated our original motivation; we believe this because the reason any particular system favored expressivity or learnability was often the result of a series of nuanced and intertwined design choices. 
We believe this paper provides a new type of contribution to the visualization community~\cite{lee2019contribution}, and demonstrates the value in making collaborative progress through in-depth critical reflection.



\bibliographystyle{abbrv-doi}

\bibliography{main}
\end{document}